# Ubiquitous stripe phase and enhanced electron pairing in interfacial high-$T_c$ superconductor FeSe/BaTiO$_3$

Xuemin Fan,[1,2] Wenqiang Cui,[1,3] Yonghao Yuan,[1,2] Qi-Kun Xue,[1,2,3,4*] and Wei Li[1,2*]

[1]*State Key Laboratory of Low-Dimensional Quantum Physics, Department of Physics, Tsinghua University, Beijing 100084, China*

[2]*Frontier Science Center for Quantum Information, Beijing 100084, China*

[3]*Beijing Academy of Quantum Information Sciences, Beijing 100193, China*

[4]*Southern University of Science and Technology, Shenzhen 518055, China*

*To whom correspondence should be addressed: weili83@tsinghua.edu.cn; qkxue@mail.tsinghua.edu.cn

Monolayer FeSe grown on SrTiO$_3$ provides a new playground for high-$T_c$ superconductivity. A recent study shows the importance of stripy instability for superconductivity enhancement in FeSe/SrTiO$_3$. However, it is still under debate whether such a stripe phase is ubiquitously bound to the interfacial superconductivity. Here, we report on molecular beam epitaxy growth and low-temperature scanning tunneling microscopy study of FeSe films on BaTiO$_3$(001). We find that the stripe phase is more prominent in FeSe/BaTiO$_3$. Long-range stripes exist in bilayer and triple-layer FeSe films at 4 K and even persist up to 77 K in the case of bilayer FeSe, with enlarged superconducting gaps upon alkali-metal deposition. The results point out an intimate correlation between the interfacial superconductivity and the stripe phase.

Understanding the emergent exotic phases[1-5] is essential for high-$T_c$ superconductivity. Interfacial high-$T_c$ superconductivity has been discovered in one unit-cell (UC) FeSe film grown on SrTiO$_3$ (FeSe/STO)[6], in which the superconducting gap-closing temperature can be as high as 65 K[7-9]. Intriguing phenomena such as charge transfer[7-9], electron-phonon coupling[10], nematicity[9, 11] and stripe phase[12] interact with each other in FeSe/STO. The roles of STO substrate in superconductivity enhancement have been widely studied. STO is believed to be a charge reservoir for electron doping into 1 UC FeSe films[7-9], giving rise to electronic structure reconstruction, which is characterized by absence of band crossing at the Fermi level ($E_F$) near the Brillouin zone center of FeSe. On the other hand, interfacial electron-phonon interaction[10, 13], manifesting as replica bands, is believed to promote the superconductivity too.

As revealed by scanning tunneling microscopy (STM)[14], there exists a novel stripe phase, manifesting itself as an incommensurate unidirectional charge order, on the surface of FeSe/STO. The stripes break the rotational and translational symmetries of the FeSe lattice, leading to a smectic electronic liquid crystal phase[1, 14]. Reminiscent of the checkerboard charge order[1-3, 15-18] and the $C_2$ charge puddle[19] with 4$a$ periodicity in cuprate superconductors, the wavelength of the stripes in FeSe films is ~ 5.1$a_0$ or 2.0 nm[14], where $a$ and $a_0$ are the lattice constant of bulk cuprate and FeSe, respectively. Long-range stripe phase only appears in 2 UC FeSe/STO, while short-range stripes pinned by defects develop in thicker films[14]. The stripe phase can be suppressed by heavy electron doping in FeSe/STO, and the remaining electronic instability is proved to further boost the superconductivity[14].

The stripe phase in FeSe/STO is closely related to the enhanced electronic anisotropy (nematicity) induced by lattice expansion[12, 14]. However, the existence of a small amount of dopants in 2 UC FeSe intertwines the charge and lattice degrees of freedom and complicates formation of the stripe phase[9, 14]. It remains elusive whether the stripe phase is ubiquitous in high-$T_c$ superconducting FeSe films. In this study, we address this problem by changing the STO substrate to BaTiO$_3$ (BTO) as schematically shown in Fig. 1(a) and 1(b), where enhanced superconductivity with gap-closing temperature of 75 K is recently observed[13].

FeSe thin films were prepared by molecular beam epitaxy (MBE) growth method. The Nb doped (0.15% wt) BTO(001) substrates were degassed in ultra-high vacuum chamber (base pressure is better than $5 \times 10^{-10}$ Torr) at 500 °C for several hours and subsequently annealed at 1050 °C for 20 min to obtain clean surfaces. High purity Fe (99.995%) and Se (99.9999%) sources were co-evaporated by two Knudsen cells to grow FeSe films. During the growth, the substrates were kept at 430 °C by applying DC current. The as-grown samples were annealed at 430 °C for one hour to improve the sample quality. The Rb deposition was performed *in situ* by using a rubidium dispenser (SAES Getters). *In-situ* STM measurements were performed at 4.2 K in a commercial STM (Unisoku). A polycrystalline PtIr STM tip was calibrated on Ag island before STM experiments. STS data were taken by a standard lock-in method. The feedback loop is disrupted during data acquisition with the frequency of oscillation signal of 973.0 Hz.

An atomically flat surface with less defects is critical for the growth of monolayer FeSe film, but this is difficult to be achieved in BTO. Compared to SrTiO$_3$ substrate, the optimal annealing temperature window for BaTiO$_3$ treatment is much narrower, which requires more precise temperature monitor and control during the experiment. With improved surface condition of substrate, the 1UC FeSe thin film is achieved in our experiment. Figure 1(c) displays a STM topographic image of 1 UC FeSe/BTO. The step height in the STM image is ~ 4.0 Å, corresponding to one UC thickness of BTO along the *c* axis. Typical tunneling spectrum is shown in Fig. 1(d), indicating that 1UC FeSe films grown on STO[20] and BTO substrates have similar electronic structure. Thus, BTO may be also a charge reservoir for 1 UC FeSe. The morphology of the BTO substrate and the grown 1UC FeSe films is not as smooth as that of STO[6], as evidenced by the dark and bright regions in Fig. 1(e). The 1 UC film shows a Se-terminated tetragonal lattice with the extended lattice constant of 4.0 Å [Fig. 1(f)]. The tunneling spectra [Fig. 1(g)] taken along the white arrowed dashed line in Fig. 1(e) exhibit clear superconducting gaps near $E_F$ with the gap sizes ranging from 9 to 13 meV. The fluctuation of the gap sizes is attributed to the inhomogeneity of BTO surface. For the same reason, the observed superconducting gaps here are not as large as that in 1 UC FeSe/STO and the treatment of BTO needs to be improved in future studies on its interfacial superconductivity. Nevertheless, for the study of stripe phase in multilayer FeSe/BTO, 1 UC FeSe/BTO could be a good enough substrate [Fig. 2(a)].

Similar to 2 UC FeSe films on STO[14], 2 UC FeSe films on BTO is not superconducting [Fig. 2(b)] either. Differential conductance (d$I$/d$V$) mapping on a 2 UC film [Fig. 2(c)] reveals pronounced long-range stripe patterns, as well as twin domains and the domain boundaries. The orientations of stripes in two adjacent domains are perpendicular to each other [denoted by white arrows in Fig. 2(c)]. Stripe patterns with Se-Se lattice are displayed in the inset of Fig. 2(c), indicating that the stripes are along the diagonal direction of the Se-Se lattice.

The stripe phase is a kind of static smectic charge order. Figure 2(d) (upper panel) shows the energy-dependent profile of the stripes, extracted from a series of differential conductance mappings at the same location with various energies. The stripes are of higher intensity within the energy ranges of -70 meV to -180 meV and 100 meV to 200 meV [upper panel of Fig. 2(d)], while its periodicity is not sensitive to the energy [lower panel of Fig. 2(d)]. Similar to that in FeSe/STO, the periodicities of the stripes fluctuate with the locations, leading to an average wavelength of 2.25 nm. We summarize the stripe periodicities at different locations in FeSe/BTO and FeSe/STO [Fig. 2(e) and (f)], respectively. The stripes are generally longer (from 1.8 nm to 3.1 nm) in FeSe/BTO [Fig. 2(e)], while the stripes in the latter range from 1.7 nm to 2.6 nm [Fig. 2(f)]. Since the lattice constant (4.0 Å) of BTO is larger than that (3.9 Å) of STO, the tensile strain may play important role for the formation of the stripes in FeSe.

The long-range stripes survive in the vicinity of domain walls on 3 UC FeSe films on BTO, marked by the red arrows in differential conductance mapping [Fig. 3(a)], while they are absent in 3 UC FeSe films on STO[14]. Although the fluctuation of the stripe wavelength in 3 UC FeSe/BTO is quite large (from 1.4 nm to 2.8 nm) [Fig. 3(b)], the thickness-dependence is similar in the two systems. In 3 UC FeSe/BTO, the intensity of the long-range stripes becomes lower [see the comparison of 2 UC and 3 UC FeSe in Fig. 3(c)]. Subsequently, in 4 UC FeSe/BTO, long-range stripes disappear and only short-range stripes can be observed near defects [Fig. 3(d)], similar to that in 3 UC FeSe/STO[14].

Temperature evolution of the stripes further reveals the enhanced stripe phases in FeSe/BTO. As shown in differential conductance mappings taken at 77 K, obvious stripe patterns and twin domains are observed in 2 UC FeSe/BTO [Fig. 3(e) and (f)]. The result here presents sharp contrast to that in 2 UC FeSe/STO, in which the stripy instability is not able to form a long-range order at 77 K. In 3 UC FeSe/BTO, the nematic domain walls are visible but the stripes are absent [Fig. 3(g) and (h)], consistent with the scenario that the stripe phase develops beneath the nematicity[12].

Similar to those in FeSe/STO[14], the stripe phases are also locally suppressed by Rb deposition in 2 UC FeSe/BTO, which leads to a dramatic phase decoherence of the stripes [Fig. 4(a)]. As a result, the stripe area ratio in 2 UC FeSe films gradually decreases with increased Rb coverage and the superconductivity starts to emerge after the stripes disappear completely [Fig. 4(b)]. Fig. 4(c-d) summarize the doping dependence of typical d$I$/d$V$ spectra in 2 UC and 3 UC FeSe thin films with most decent superconducting gap features, including symmetric (with $E_F$) superconducting gaps and clear coherent peaks. Moreover, the averaged superconducting gap size evolution in FeSe/BTO with Rb deposition is shown in Fig. 4(e). They both present dome-like features with fluctuations because of sample inhomogeneity. The corresponding superconducting gap sizes at optimum doping are 14.7 meV and 13.6 meV for 2 UC and 3 UC Rb-doped FeSe/BTO, respectively [Fig. 4(f)]. They are larger than the gap ~ 12.1 meV in 2 UC Rb-doped FeSe/STO (with weaker stripes)[14]. The enlarged gap size here indicates that the enhanced stripe phase corresponds to stronger potential for electron pairing. In addition, the superconducting gap-closing temperature in 1 UC FeSe/BTO is higher than that in 1 UC FeSe/STO (75 K vs. 65 K)[8, 13]. We therefore conclude that superconductivity and stripe phase are both enhanced in FeSe/BTO.

We plot the phase diagrams of the FeSe/BTO and FeSe/STO [Fig. 5(a) and (b)] as a function of temperature and film thickness. For 1 UC FeSe, high-$T_c$ superconductivity exists in both systems. For thicker FeSe films, nematic and stripe phase emerge beside the superconducting phase, and stripe phase develops as the strength of the nematicity becomes stronger at low temperature or with decreased film thickness. For FeSe films thicker than 3 UC, short-range stripes appear in the vicinity of the defects, where further symmetry breaking occurs[12].

By comparison of two systems, we can decouple the charge and lattice degrees of freedom for their effects on the stripes. The key variable between FeSe/BTO and FeSe/STO is the increased tensile strain [see the insets of Fig. 5(a) and (b)], which in turn leads to stronger electronic anisotropy and correlation. In addition, consistent with the previous studies[12, 14], the itinerant Fermi surface nesting picture is not suitable to the observed stripes here. Therefore, the current results indicate that the lattice expansion and the consequent enhanced electronic anisotropy and correlation are the main causes of the stripe phase, rather than the effect of charge doping, which actually suppresses the electronic anisotropy instead[14]. Moreover, charge transfer from substrate to 3 UC FeSe film is negligible[9] but the long-range stripe pattern can still exist in 3 UC FeSe/BTO near the domain boundaries [Fig. 3(a) and (b)], where larger lattice expansion is expected[21]. This observation further highlights the significant role of electronic correlation for the formation of stripes.

In summary, we have successfully demonstrated the positive correlation between the stripes and superconductivity in FeSe/oxides. The strongly correlated interactions play significant roles not only for the emergence of the stripes, but also for the superconductivity enhancement. The following approaches are considered for the future studies: (1) Realizing higher $T_c$ in FeSe thin films grown on oxides with larger lattice constant, for example, $PrScO_3$ and $BaSnO_3$. But the interfacial phonon coupling and charge transfer from the substrates should be reconsidered, and the work functions are different as well. (2) Investigating the superconducting properties in FeSe single crystal with applying tensile strain and gate/surface doping. By doing this, the lattice constant and doping level can be tuned independently and continuously, so that the whole phase diagram of FeSe can be mapped out. We believe our findings on the interactions between the stripe phase, charge order, nematicity and superconductivity, reminiscent of those in iron-based[4, 22-30] and cuprate[1-3, 17, 31-34] superconductors, shed important light on understanding the mystery of high-$T_c$ superconductivity.

We thank Yayu Wang, Hong Yao and Yan Zhang for inspiring discussions. The experiments were supported by the National Science Foundation (No. 51788104, No. 11427903, No. 11674191), the Initiative Research Projects of Tsinghua University (No. 20211080075) and the Beijing Advanced Innovation Center for Future Chip (ICFC). W. Li was also supported by Beijing Young Talents Plan and the National Thousand-Young-Talents Program.

**Figures and Captions**

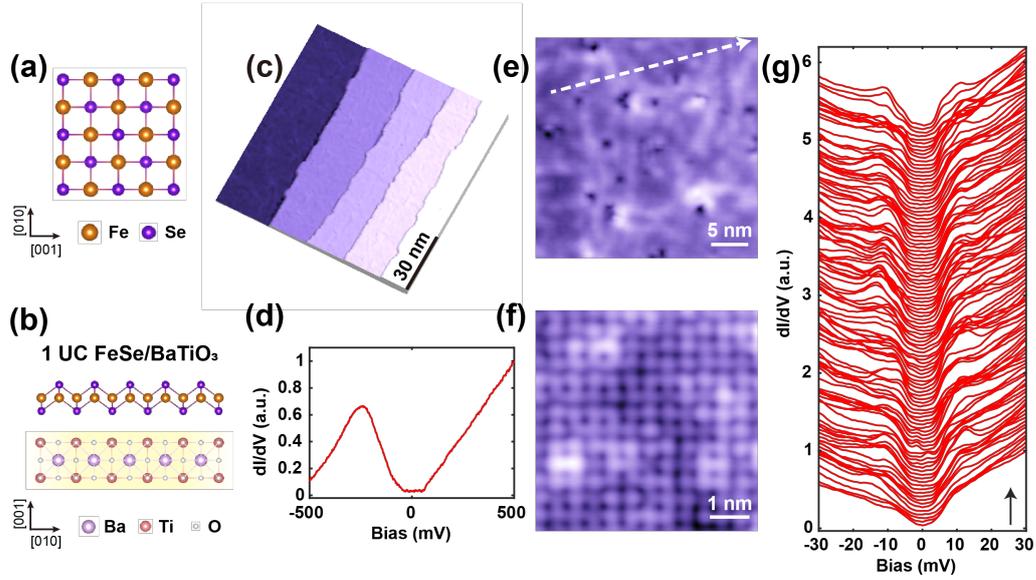

FIG. 1. Superconductivity in 1 UC FeSe/BTO. (a) Lattice structure of FeSe. (b) Schematic of 1 UC FeSe epitaxially grown on BTO. (c) STM topographic image of a 1 UC FeSe/BTO (100 nm × 100 nm; set point, $V_s$ = 1.0 V, $I_t$ = 20 pA). The BTO is fully covered and the step height is 4.0 Å. (d) Typical d$I$/d$V$ spectrum of 1 UC FeSe (set point, $V_s$ = 500 mV, $I_t$ = 200 pA). (e) STM topographic image of 1 UC FeSe (30 nm × 30 nm; set point, $V_s$ = 500 mV, $I_t$ = 20 pA), in which the dark and bright parts originate from the inhomogeneity of BTO substrates. (f) Atomically resolved topographic image of 1 UC FeSe (5 nm × 5 nm; set point, $V_s$ = 60 mV, $I_t$ = 100 pA). (g) A series of spectra taken along the white arrow in (e) (set point, $V_s$ = 60 mV, $I_t$ = 200 pA). The spectra of 1 UC FeSe show clear superconducting gaps.

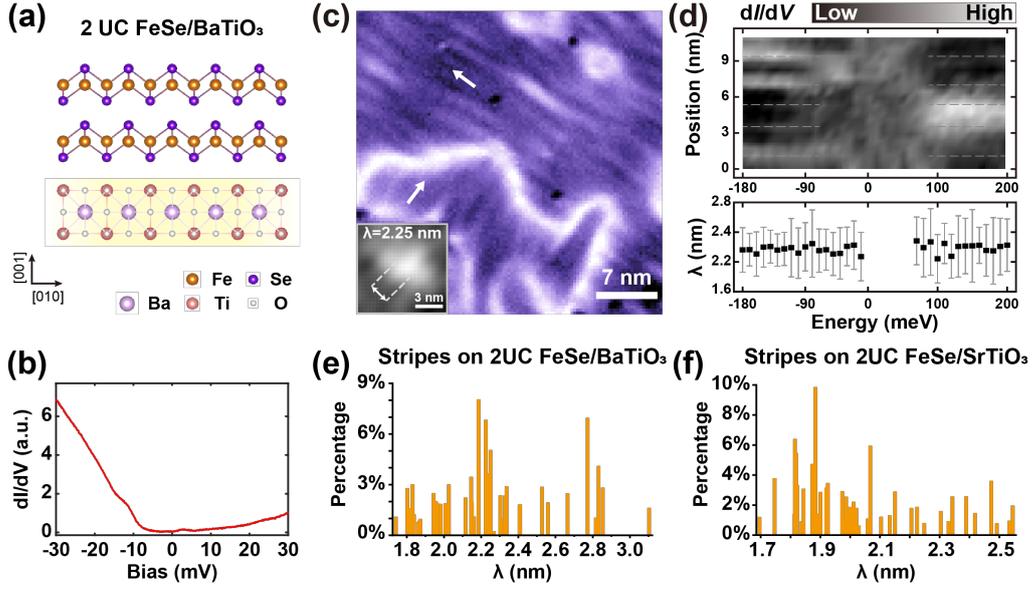

FIG. 2. Stripe phase in 2 UC FeSe/BTO. (a) Schematic of 2 UC FeSe/BTO. (b) d$I$/d$V$ spectrum of 2 UC FeSe (set point, $V_s$ = 60 mV, $I_t$ = 200 pA). 2 UC FeSe is not superconducting. (c) A STM d$I$/d$V$ mapping of 2 UC FeSe/BTO at -120 mV (35 nm × 35 nm; set point, $V_s$ = 150 mV, $I_t$ = 20 pA). White arrows denote the orientation of stripes in different smectic domains. Inset: atomically resolved topographic image of 2 UC FeSe shows the periodicity ~ 2.25 nm and orientation of the stripes. (d) Upper panel: d$I$/d$V$ line-cut of stripes as a function of energy. Lower panel: The period of the stripes as a function of energy. The periodicity is calculated by $l/n$, where $l$ is the length of the line-cut and $n$ is the number of the nodes in each line-cut. The error bar denotes the standard deviation calculated from the distance of two adjacent nodes. (e-f) The period distribution of the stripes on 2 UC FeSe/BTO and 2 UC FeSe/STO, respectively. The wavelengths are determined from the peak positions in the FFT results of the corresponding d$I$/d$V$ mappings with obvious stripes at different locations. The height of each volume is calculated by $S_i/S$, where $S_i$ is area of each d$I$/d$V$ mapping and $S$ is the total area.

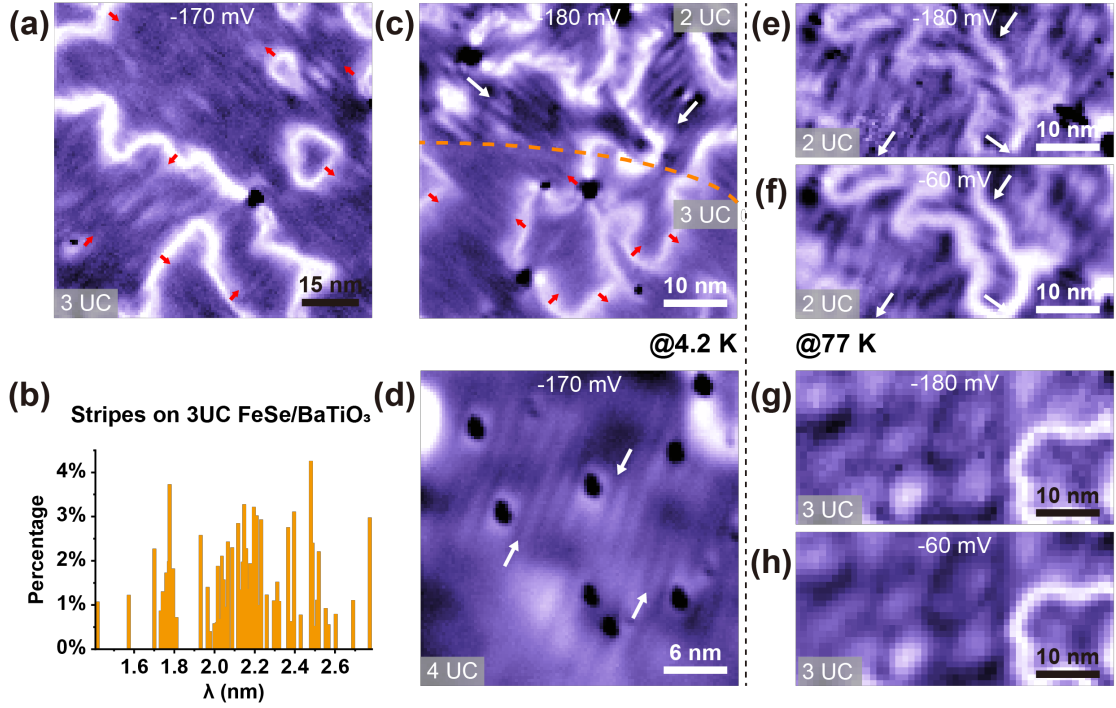

FIG. 3. Enhanced stripe phase in FeSe/BTO. (a) STM d$I$/d$V$ mapping of a 3 UC FeSe/BTO at -170 mV (60 nm × 60 nm; set point, $V_s$ = 120 mV, $I_t$ = 20 pA). Red arrows denote the stripe order in the vicinity of domain walls. (b) The period distribution of the stripes of 3 UC FeSe/BTO. The algorithm is identical to that of Fig. 2(e-f). (c) d$I$/d$V$ mapping taken on an area including 2 UC and 3 UC FeSe/BTO at -180 mV (50 nm × 50 nm; set point, $V_s$ = 120 mV, $I_t$ = 20 pA). The stripe patterns are weaker in 3 UC FeSe. (d) d$I$/d$V$ mapping of 4 UC FeSe/BTO at -170 mV (30 nm × 30 nm; set point, $V_s$ = 120 mV, $I_t$ = 20 pA). The stripes are induced by impurities with $C_2$ symmetry, indicated by white arrows. (e-f) d$I$/d$V$ mappings (50 nm × 25 nm; set point, $V_s$ = 120 mV, $I_t$ = 200 pA) of 2 UC FeSe/BTO at 77 K. The stripes and the domain walls can be clearly observed. (g-h) d$I$/d$V$ mappings (50 nm × 25 nm; set point, $V_s$ = 120 mV, $I_t$ = 200 pA) of 3 UC FeSe/BTO at 77 K. The domain walls still exist but the stripes disappear, indicating a smectic-to-nematic phase transition.

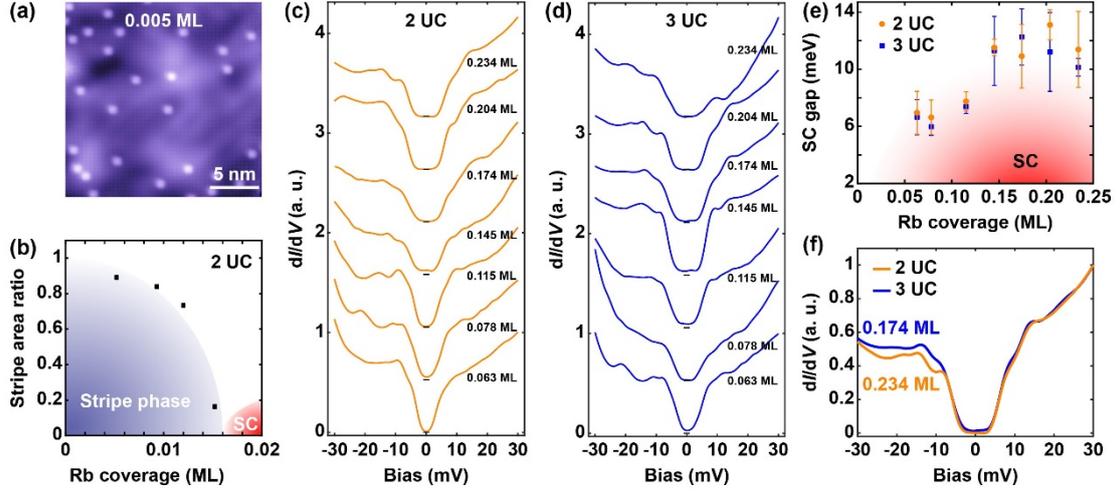

FIG. 4. Doping evolution of the stripe and superconducting phase in FeSe/BTO. (a) STM topography (20 nm × 20 nm; set point, $V_s$ = 100 mV, $I_t$ = 100 pA) of Rb-doped 2 UC FeSe/BTO with the Rb-coverage of 0.005 monolayer (ML). 1 ML is defined as 1 Rb atom per Fe site. (b) Suppression of stripe phase in 2 UC FeSe. The blue and red shaded regions denote the stripe and superconducting states, respectively. (c-d) Doping dependence of typical d$I$/d$V$ spectra taken on 2 UC and 3 UC FeSe, respectively. (e) Superconducting gap dependence on Rb coverage in 2 UC and 3 UC FeSe. The error bars denote the standard deviations of superconducting gap sizes at different Rb coverages. (f) Superconducting gaps of Rb-doped 2 UC and 3 UC FeSe/BTO with optimum doping (set point, $V_s$ = 60 mV, $I_t$ = 120 pA).

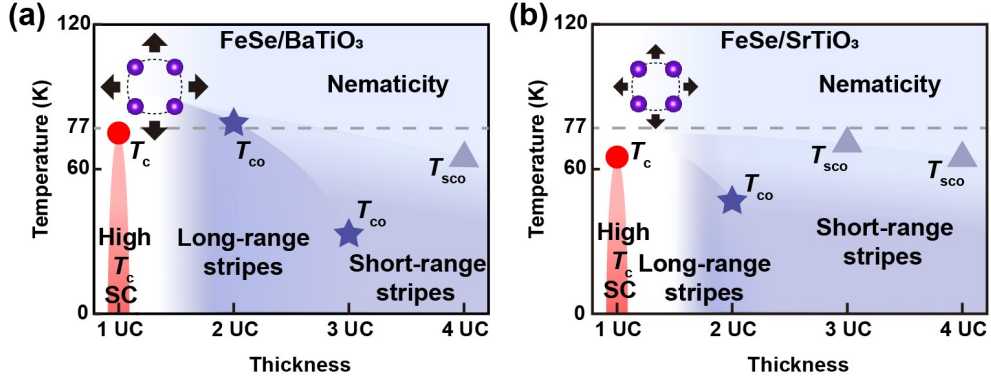

FIG. 5. Phase diagrams of FeSe/BTO and FeSe/STO. (a) Phase diagram of FeSe/BTO as a function of temperature and thickness. 1 UC FeSe is a high-$T_c$ superconductor. In 2 UC and 3 UC FeSe, long-range stripe order is established and their critical temperatures ($T_{CO}$) are denoted by blue pentagrams. The stripe phase can survive above 77 K in 2 UC FeSe/BTO. In 4 UC FeSe, short-range stripes are observed and the critical temperature ($T_{SCO}$) is denoted by blue triangle. Inset: schematic of the enlarged tensile strain in FeSe/BTO. (b) Phase diagram of FeSe/STO as a function of temperature and thickness. The long-range stripe phase only exits in 2 UC and disappears in 3 UC FeSe as a result of decreased tensile strain in FeSe/STO (inset). Besides the results in the current experiment, some nematic and stripe phase transition temperatures are extracted from previous STM[12, 21] and ARPES[9] results. The superconducting gap-closing temperatures ($T_C$) denoted by red dots are obtained from ARPES results[8, 13]. As shown in the phase diagrams, the superconductivity and the stripe phase are both enhanced in FeSe/BTO.